\documentclass[a4paper]{article}
\usepackage[english]{babel}
\usepackage[utf8x]{inputenc}
\usepackage[T1]{fontenc}

\usepackage[a4paper,top=3cm,bottom=2cm,left=3cm,right=3cm,marginparwidth=1.75cm]{geometry}

\usepackage{amsmath, amssymb}
\usepackage{textcomp}
\usepackage{gensymb}
\usepackage{amsthm}
\usepackage{graphicx}
\usepackage{caption}
\usepackage{subcaption}
\usepackage{adjustbox}
\usepackage{booktabs}
\usepackage{epstopdf}
\usepackage{placeins}
\usepackage[colorlinks=true, allcolors=blue]{hyperref}
\usepackage{cite}

\usepackage{setspace}
\usepackage{pdfpages}

\title{Effects of particle roughness on the rheology and structure of capillary suspensions}
\author{Jens Allard$^1$, Sanne Burgers$^{1,\dag}$, Miriam Candelaria Rodr\'{i}guez Gonz\'{a}lez$^2$, \\
Yanshen Zhu$^1$, Steven De Feyter$^2$, Erin Koos$^{1,\ast}$}
 \date{ \small
 $^1$ {KU Leuven, Department of Chemical Engineering, Celestijnenlaan 200J, 3001 Leuven, Belgium} \\
$^2$ {KU Leuven, Department of Chemistry, Celestijnenlaan 200F, 3001 Leuven, Belgium}  \\
 $^{\ast}$ E-mail: erin.koos@kuleuven.be \\
 $^\dag$ Present address: Rijksweg 12, 2870 Puurs, Belgium \\}

\begin{document}

%\tracingall
\maketitle

\begin{abstract}
We show that particle roughness leads to changes in the number, shape and resulting capillary force of liquid bridges in capillary suspensions. We created fluorescently labeled, raspberry-like particles with varying roughness by electrostatically adsorbing silica nanoparticles with sizes between 40 nm and 250 nm on silica microparticles. Rougher particles require more liquid to fill the surface asperities before they form pendular bridges, resulting in smaller and weaker bridges. In a system where the effective bridge volume is adjusted, higher particle roughness leads to less clustered networks, which show a higher yield strain for a matching storage modulus compared to the smooth particle networks. This finding suggests that the particle-particle frictional contacts also affects the strength of capillary suspensions. Using asymptotically nonlinear oscillatory rheology, we corroborate the non-cubical power law scaling of the third harmonic in the shear stress response that results from both Hertzian contacts and friction between particles connected by capillary bridges. We demonstrate that the repulsive Hertzian contact parameter $A$ is sensitive to the liquid bridge strength and that roughness appears to shift the relative scaling of the power law exponents from adhesive-controlled friction to load-controlled friction.  
\end{abstract}

%%%MAIN TEXT%%%%
\section{Introduction}
Particle surface roughness has been proven to significantly affect the behavior of many systems, like suspensions, Pickering emulsions, and wet granular media.~\cite{More2020,Hu2020,Butt2009} Surface roughness alters the frictional contacts in dense particle systems and plays a role in the wetting interactions between liquids and solids by changing the three phase contact angle, for example for a liquid spreading on a surface or a particle adsorbing at or desorbing from a liquid-liquid interface. Thus, the ability to engineer surface roughness is of major interest in the tuning of flow behavior of particle suspensions,~\cite{Hsu2018,Hsu2021} the stabilization of Pickering emulsions~\cite{San-miguel2012,Zanini2017} or the development of self-cleaning surfaces.~\cite{Geyer2020}

A relatively new class of materials are capillary suspensions, three-phase mixtures of solids and two liquids where the secondary liquid is only present as a few percent. The particles are connected by liquid menisci of the secondary liquid to form a sample-spanning network in the bulk liquid. This network formation drastically increases the viscosity and either forms or increases the yield stress compared to the binary suspension without secondary liquid, often creating a gel-like rheological response.~\cite{Koos2011} Depending on the three-phase contact angle, a morphological distinction is often made between pendular state and capillary state capillary suspensions for a preferentially ($\theta<90 \degree$) or a non-preferentially ($\theta>90 \degree$) wetting secondary liquid, respectively.~\cite{Bossler2016} These capillary suspensions can be applied in coatings,~\cite{Fischer2021} low fat-foods,~\cite{Hoffmann2014,Wollgarten2016} printable electronics,~\cite{Schneider2016} or as precursors of ceramic materials.~\cite{Dittmann2013,Dittmann2014,Dittmann2016} It is important to note that capillary suspensions are not in thermodynamic equilibrium and, as a consequence, the structure and rheological properties are sensitive to the preparation method and shear history of the material.~\cite{Bossler2017,Koos2014,Hao2021} With increasing secondary liquid volume, an increasing number of liquid bridges are formed between particles and their relative volume increases. The bulk mechanical properties, such as storage modulus or yield stress, increase accordingly.~\cite{Bindgen2020,Huprikar2020,Domenech2015} In the pendular state, this trend continues until liquid menisci of neighboring particles start to merge together, which denotes the maximum strength of the capillary suspension and this is called the funicular state. Increasing the liquid volume further decreases the strength due to particle aggregation.~\cite{Bindgen2020,Bindgen2021,Roh2017}

Particle roughness has several key implications for capillary suspensions: it will affect the shape, strength and number of the liquid bridges, as well as the frictional particle contacts themselves. The formation and strength of liquid menisci between two rough particles has been studied in wet granular media.~\cite{Halsey1998,Butt2008,Kohonen2004} Similar to capillary suspensions, the addition of a small amount of secondary liquid to a granular pile drastically increases the yielding pressure due to the formation of capillary bridges.~\cite{Fournier2005} Increasing the amount of secondary liquid will increase the number and volume of the bridges. For rough particles, the surface asperities first need to be (partially) filled with secondary liquid before a bridge can be formed. Therefore, the capillary force is initially low when a bridge forms from a single asperity, increases with secondary fluid as the number of these asperity bridges increases, and then continues to increase with secondary liquid volume until a single bridge is formed that extends laterally over the particle surface. Halsey and Levine divided the capillary force versus secondary liquid effect for wet granular media in three regimes, namely the asperity, roughness and spherical regimes.~\cite{Halsey1998} The asperities trap a fraction of the secondary liquid into a ``film" surrounding the particles. With increasing roughness, the volume of the fluid trapped in this film increases, thereby reducing the size and number of the bridges between adjacent particles. Roughness would, thus, be expected to decrease the capillary force,~\cite{Butt2009} a result which was also obtained for the capillary force between parallel plates covered with sandpaper.~\cite{Nguyen2021} In some specific cases, however, the capillary force may actually increase with a larger surface roughness. If the bridges are able to divide and migrate to local gap height minimums, which is possible if the contact line does not get pinned, the total capillary force increases with the roughness amplitude and forms a resistance to shear.~\cite{Butler2022} 

Even in the case where the capillary suspensions only have bridges between particles and not asperities (the spherical regime), the roughness will still influence the capillary force. The three-phase contact angle and the particle or local curvature are crucial parameters in all systems involving wetting, including capillary suspensions due to their dependence on the shape and strength of the bridges. In general, surface roughness moves the contact angle away from neutral wetting, i.e. hydrophilic particles become more hydrophilic and vice versa.~\cite{Ballard2019} However, when the contact line gets pinned, the contact angle can adopt a variety of metastable positions, which increases the contact angle hysteresis.~\cite{Marmur1994} This pinning effect has been successfully used to increase the stabilization of Pickering emulsions~\cite{San-miguel2012,Zanini2017} and was also observed for the liquid bridges in capillary suspensions.~\cite{Bossler2016} To take the effect of particle roughness on the static contact angle into account, the theories of Wenzel and Cassie-Baxter were developed. The former describes a homogeneous wetting of the asperities where the contact angle change is given by:
\begin{equation}
   \cos(\theta_\mathrm{{rough}}) = r_w \cos(\theta_\mathrm{{smooth}})
    \label{eq: Yan contact angle}
\end{equation}
where $\theta$ is the contact angle of the bridging fluid and $r_w$ is the ratio between the total surface area and projected surface area.~\cite{Yan2007} In the Cassie-Baxter case, the bulk liquid is trapped underneath the wetted areas and the contact angle change depends on the area fraction of wetting liquid-solid contact. San-Miguel and Behrens showed that the transition from Wenzel to Cassie-Baxter wetting decreases the contact angle hysteresis, and the resulting pinning force, and required desorption energy of microparticles at the surface of Pickering droplets, leading to a decreased stability of the emulsions.~\cite{San-miguel2012} This pinning effect due to contact angle hysteresis was also recently shown to enhance the capillary torque of a capillary bridge between a particle and a liquid-liquid interface.~\cite{Naga2021,naga2021force} Therefore, Wenzel wetting might be preferred over Cassie-Baxter wetting to create the most stable capillary suspensions. However, both theories describe an equilibrium contact angle and do not take into account the local surface roughness or the possibility of metastable contact angle positions.

Finally, the particle roughness can influence the breaking of capillary bridges. When a capillary suspension is sheared outside of the linear viscoelastic regime, the particle network yields and some bridges are stretched, compressed or broken. These explicit bonds between particles make the capillary suspension networks very different from other attractive particle networks, for example originating from depletion interactions, and result in rigid body movement of the particle flocs.~\cite{Bindgen2021} The rupture dynamics experiments of Zanini et al.\@ showed that a pinned contact line reduces the rupture force and rupture distance when stretching the liquid meniscus between a rough particle and a liquid-liquid interface compared to a sliding contact line when using a smooth particle.~\cite{Zanini2018} However, pinned contact lines and an increased hysteresis may also increase the rigid body movement in these systems.

Even if the influence of roughness on the capillary bridges is neglected, it still influences the resistance to shearing via friction. In regular suspensions, frictional particle contacts and the friction coefficient can be linked to changes in the relative viscosity~\cite{More2020}, as well as to the shear thinning~\cite{Chatte2018,Lobry2019,Gallier2014} and shear thickening~\cite{Hsu2018,Hsu2021} behavior. The shear thickening experiments, for example, show that the jamming transition occurs at a much lower particle volume fraction in rough particle systems due to the interlocking asperities.~\cite{Hsu2018} In capillary suspensions, particles are explicitly pulled into contact by the capillary bridges, which introduces a Hertzian repulsion term in addition to the capillary force term to describe the force between two particles connected by a liquid bridge.~\cite{Butt2010} Natalia et al.\@ found that this combination of Hertzian contact and capillary force explains the non-integer power law scaling of the third harmonic elastic stress in the medium amplitude oscillatory shear regime.~\cite{Natalia2022} Typically, the third harmonic elastic and viscous stresses $\sigma_3$ are expected to scale with an exponent of $m=3$ with the applied strain, $\sigma_3 \propto \gamma_0^3$, but for capillary suspensions, as well as for some other particulate systems, this power law scaling $m$ was shown to be noncubic.~\cite{Natalia2020} In that work, Natalia et al.\@ linked the scaling of the third harmonic viscous stress to particle-particle friction. 

All of these findings suggest that changing the particle roughness in capillary suspensions, which combine the themes of wetting and particles in contact, likely affects their microstructure and bulk rheological properties in many non-trivial ways. In the present work, we examine this effect using raspberry particles with well-controlled roughness, synthesized via the adsorption of nanoparticles onto the microparticle surface, secured with a silica layer. We describe the influence of roughness on capillary suspensions with both equal amounts of secondary liquid and systems where the volume of the secondary fluid is adjusted to match the strength. These capillary suspensions are described both in terms of their mechanical properties using bulk rheological tools and structure using confocal microscopy. Finally, using medium amplitude oscillatory shear, we can investigate the nature of the Hertzian and frictional contacts. 

\section{Materials and methods}
Silica particles (SOLAD non-porous PNPP3.0NAR, Glantreo) with an average diameter of 3~$\mu$m and polydispersity $< 5\%$ were used to prepare rough particle and smooth particle capillary suspensions. The silica particles were fluorescently labeled using rhodamine B isothiocyanate (RBITC, Sigma-Aldrich) or fluorescein isothiocyanate (FITC, Sigma-Aldrich) via a modified St\"{o}ber synthesis, after which the dye is covalently bonded to the particle surface. FITC was only used due to a shortage of RBITC, as it is more prone to photobleaching during the 3D-image scanning. The density of the smooth particles was found to be 2.15 $\pm$ 0.11~g/ml, determined using a buoyancy measurement (Sartorius Density Determination Kit YDK01) in air and in ethanol. 

\subsection{Rough particle synthesis}
To prepare particles with varying roughness, the synthesis method for all-silica raspberry particles based on electrostatic adsorption of nanoparticles on microparticles by Hsu et al.\@ was used.~\cite{Hsu2018} 
Ethanol (99.8 \%, Fischer Chemical), ammonium hydroxide (28-30 \%, Merck) and tetraethyl orthosilicate (TEOS, 98 \%, Acros Organics) were used in a St\"{o}ber reaction to synthesize nanoparticles of approximately 40 nm, 100 nm and 250 nm in ethanol. The particle size was verified using a commercial 3D dynamic light scattering device (3D DLS, LS Instruments). The scattering angle was varied from 30{\degree} to 150{\degree} at 15{\degree} intervals. The detected nanoparticle diameters, synthesis parameters and mixing times can be found in Table~\ref{tbl:Nanoparticle recipes}. 
\begin{table}[tbp]
\small
  \caption{\ Number averaged hydrodynamic diameter detected using 3D dynamic light scattering and nanoparticle synthesis components}
  \label{tbl:Nanoparticle recipes}
  \begin{tabular}{ccccc}
    \toprule
    Size    & Ethanol   & NH$_4$OH  & TEOS  & Mixing time \\
    
    [nm]    & [ml]      & [ml]      & [ml]  & [hours] \\
    \midrule
    41 $\pm$ 17 & 92 & 3.8 & 4.5 & 2 \\
    103 $\pm$ 28 & 90 & 5.6 & 7.6 & 2 \\
    248 $\pm$ 67 & 88 & 7.6 & 4.5 & 24 \\
    \bottomrule
  \end{tabular}
\end{table}
At the end of the mixing time, a rotavapor (Buchi Rotavapor R215) was used at 55~{\degree}C, 20~rpm and 200~mbar for 5~minutes to evaporate most of the ammonia hydroxide and stop the growing process of the St\"{o}ber reaction.

Adsorbing the nanoparticles onto the microparticle surface was done in three steps. First, dyed silica microparticles were dispersed in a 1~vol\% solution of poly-diallyldimethylammonium chloride (poly-DADMAC, Sigma-Aldrich), which is a cationic polyelectrolyte, in Milli-Q water and stirred for 24 hours. The microparticles are centrifuged and washed with Milli-Q water 3 times at 1500 rpm using an Avanti J-30I centrifuge to remove excess poly-DADMAC. Second, the coated microparticles are dispersed in 100 ml Milli-Q water and 60 ml of the nanoparticle solution is added to this suspension. The resulting mixture is stirred for 24 hours, after which another centrifuge step is performed to extract the microparticles. Finally, a modified St\"{o}ber reaction is used to form a silica layer to improve adhesion between nano- and microparticles and to tune the final roughness. For this reaction, the microparticles are added to a mixture of 50 ml Milli-Q water, 87 ml ethanol and 15 ml ammonium hydroxide which is continuously stirred. To this mixture, 3 cycles of 2.67 ml of a 5 vol\% TEOS in ethanol solution was added by adding 167 $\mu$l every 30 seconds for 8 minutes. After each cycle, a growing period of 25 minutes was used during which no TEOS is added. After the final cycle, the solution is again centrifuged and the particles are dried overnight at 60 {\degree}C with the last hour under vacuum.

\subsection{Atomic force microscopy}
\begin{figure}[tbp]
    \centering
    \includegraphics[width=0.75\textwidth]{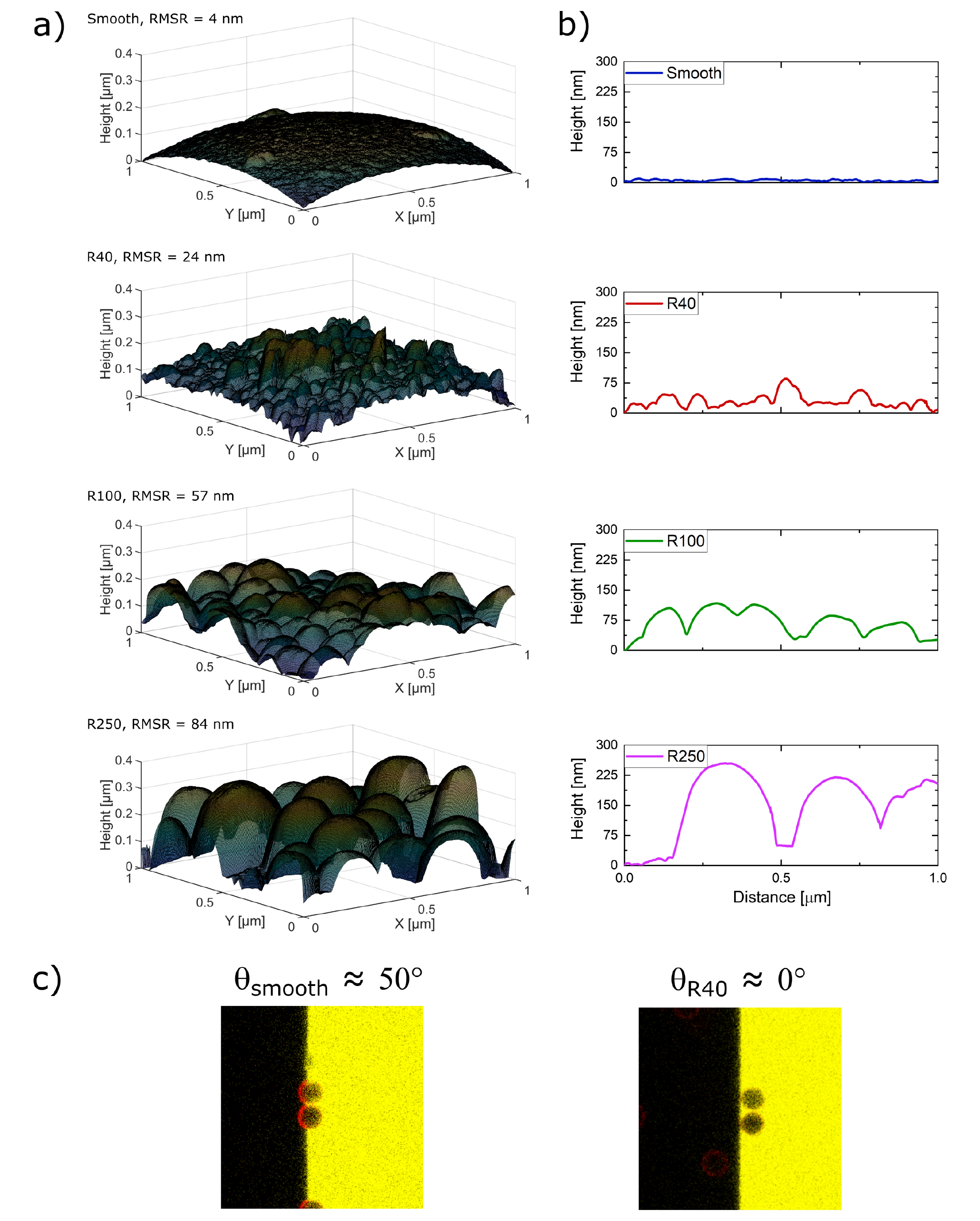}
    \caption{a) Surface profiles of silica particles without and with adsorbed nanoparticles of different size. The root mean squared roughness (RMSR) is reported above each profile. b) Curvature-corrected asperity height profile for the different particle types c) Contact angle measurement of smooth and rough R40 particles (red rings) adsorbed at a liquid-liquid interface (secondary liquid in yellow, bulk liquid in black). }
    \label{fig: Surface profiles and contact angle}
\end{figure}

Atomic force microscopy (AFM) was used to characterize the roughness of the particles. Approximately 50 mg of rough particles was dispersed in 1 ml of Milli-Q water rendering a 2 vol\% suspension, after which 100 µl was evaporated on a mica surface. AFM imaging was performed using a Multimode 8 (Bruker) with a Nanoscope V controller in tapping mode. A OMCL-AC160TS-R3 probe with a spring constant of approximately 20 N/m and a resonance frequency of approximately 300 kHz was used at a scan rate of 0.5 -- 1 Hz. The obtained images were analyzed using the open source Gwyddion software~\cite{NecasKlapetek2012} to extract the RMS roughness using images of 1 µm by 1 µm after correcting the profiles for the curvature of the particle. The surface profiles were plotted in Matlab. Microparticles with different roughness were obtained by tuning the size of the adsorbed nanoparticles. In the remainder of the text, the rough particles with adsorbed 40 nm nanoparticles are referred to as R40, and likewise for R100 and R250. The smooth and rough microparticle surface profiles obtained using AFM (without any correction for the particle radii) are shown in Figure~\ref{fig: Surface profiles and contact angle}a. 
The particles are uniformly covered by nanoparticles and with increasing nanoparticle size, the curvature-corrected final asperity height and root mean squared (RMS) roughness increase accordingly. The RMS roughness ranges from 4 nm for the smooth particles to 84 nm for the R250 particles. Figure~\ref{fig: Surface profiles and contact angle}b illustrates the increased asperity height and asperity-to-asperity distance with increasing RMS after the curvature correction. Moreover, the particle radii detected by the particle detection program also increase with increasing asperity height, which gives another indication that the nanoparticle adsorption procedure was successful. 

\subsection{Capillary suspension preparation}
Capillary suspensions were prepared using a refractive index-matched system ($n = 1.455$) with the dyed silica particles as the solid phase, a mixture of 83.8 vol\%  1,2-cyclohexane dicarboxylic acid diisononyl ester (Hexamoll DINCH, BASF) and 16.2 vol\% n-dodecane ($>$99 \%, Sigma-Aldrich) as the bulk phase, and a mixture of 16.4 vol\% deionized water and 83.6 vol\% glycerol ($>$99.5 \%, Carl Roth) as the secondary phase. The interfacial tension between these two liquids was determined to be 21.5 $\pm$ 0.3 mN/m by Bossler et al.~\cite{Bossler2016} The bulk and secondary liquid mixtures have a density of 0.91 g/ml and 1.23 g/ml, respectively. The aqueous glycerol was dyed using a small amount of water-soluble Promofluor-488 or Promofluor-780 premium carboxylic acid (PromoKine), while the bulk fluid remained undyed. The volume fraction of particles was kept constant at 20 vol\% for all capillary suspensions, while the secondary liquid was varied between 0.1\% and 5\% depending on the particle roughness. Capillary suspensions were prepared by 2-step ultrasonic mixing (Digital Sonifier model 450, Branson Ultrasonics corporation) using a 12.5 mm diameter ultrasonic horn placed inside the sample. First, both liquids were emulsified for 30 seconds at 30 \% power, thereby creating small drops of secondary liquid. Next, the particles were added and the mixture was sonicated three more times at 10 \% for 10 seconds. This two-step mixing procedure allows for a better droplet breakup and results in a more homogeneous bridge size distribution compared to the method of adding the particles first.~\cite{Bossler2017,Domenech2015} In between ultrasonication steps, the mixture is stirred with a spatula to improve the homogeneity of the sample. Suspensions without secondary liquid were prepared using the same method.  

\subsection{Rheological measurements}
Oscillatory measurements were performed on a stress-controlled MCR702 rheometer (Anton Paar) using an 8 mm parallel plate geometry. The rheometer was operated using a double motor setup for improved sensitivity. The viscoelastic moduli were confirmed to be frequency-independent for angular frequencies between 0.1 rad/s and 100 rad/s, shown in supplemental Figure S1. Therefore, all rheological measurements were performed at an angular frequency of 10 rad/s. Each measurement was performed in triplicate.   
For the medium amplitude oscillatory shear (MAOS) experiments, the protocol described by Natalia et al.\@ was performed on a strain-controlled ARES G2 rheometer (TA Instruments) using an 8 mm parallel plate geometry.~\cite{Natalia2020} Three forward and three reverse amplitude sweeps up to a maximum strain were performed, after which the procedure is repeated with a higher maximum strain, e.g. 6 sweeps until $\gamma_{max} = 0.1\%$, 6 sweeps until $\gamma_{max} = 1\%$, and so on.   

All measurements were performed at room temperature at a measurement gap of 1 mm. To prevent a change of the sample structure during loading on the rheometer, a slow gap setting profile was used and overfilling was avoided. Applying a high normal force to the sample compresses the particle clusters, pushes some bulk liquid out of the measuring gap and causes an upward shift in the viscoelastic moduli due to the increased effective solid volume fraction, which is illustrated in supplemental Figure S2. Therefore, the normal force during loading was monitored and measurements with normal force greater than 0.05 N, and hence shifted moduli, were discarded to ensure comparability between different samples.
The main reason for using the 8 mm plate was the limited available sample volumes owing to the time-consuming and labor-intensive rough particle synthesis procedure. Afterwards, some tests were repeated with a 25 mm parallel plate, covered with sandpaper to limit slip, to confirm the findings obtained with the 8 mm plate. 

\subsection{Confocal microscopy and image processing}
A Leica TCS SP8 inverted confocal laser scanning microscope was used to take 3D image stacks with a size of 82 $\mu$m $\times$ 82 $\mu$m $\times$ 100 $\mu$m at a resolution of 1024 pixels $\times$ 1024 pixels $\times$ 300 pixels. A 63x magnification glycerol immersion objective (n = 1.46) with a numerical aperture of 1.3 was used for imaging. Two solid-state lasers were used to illuminate the sample depending on the dye combination of secondary liquid and particles: 488 nm/552 nm to excite Promofluor 488/rhodamine B isothiocyanate and 638 nm/488 nm to excite Promofluor 780/fluorescein isothiocyanate. For each capillary suspension, five image stacks were taken at different positions in the sample, containing between 6000 and 14000 particles per stack, which resulted in approximately 30,000 to 50,000 particles in total per sample. The particle positions and radii were determined using a custom-written 3D detection algorithm based on edge detection and Hough transform, written in IDL, described in Bindgen et al.~\cite{Bindgen2020} An illustration of the particle detection program is shown in supplemental Figure S3. Based on the particle positions, the semi-local measures of coordination number ($z$) and clustering coefficient ($c$) were determined to describe the network structure. In the same work of Bindgen et al.,\@ it was shown that the average coordination number and clustering coefficient for capillary suspension networks increase with increasing amount of secondary liquid.~\cite{Bindgen2020} The viscoelastic moduli increased accordingly as the networks changed from the pendular to the funicular state. The clustering coefficient of a particle is defined as:
\begin{equation}
    c = \frac{2e}{z(z-1)}
\end{equation}
where $e$ represents the number of connections between the neighbors of said particle and $z$ is the coordination number. Following this definition, the clustering coefficient has a value between 0 and 1, representing a more string-like or clustered network, respectively. In physically real networks, an average clustering coefficient of 1 is not realistic and a value around 0.5 is considered as highly clustered. A perhaps more intuitive implication of the clustering coefficient is that it represents the relative number of triangles in a network, which results in more stable and rigid networks for high clustering coefficients.~\cite{Bindgen2020} 
Particles are considered to be in contact if their surface-to-surface distance is lower than a threshold of 6 pixels or 0.48 $\mu$m, which is of the same order of magnitude as the resolution of the confocal microscope in the z-direction and also takes into account possible errors in the detected particle radius. The clustering coefficient was calculated from the detected particle positions using the Python package NetworkX. 

\section{Results and Discussion}
\subsection{Effective contact angle}
The contact angles of single particles were observed using confocal microscopy. A liquid-liquid interface with the bulk and secondary liquid was created by injecting both liquids in a custom-built glass microchannel.~\cite{Bossler2016} The contact angle of the particles is determined by their position at this interface, as shown in Figure \ref{fig: Surface profiles and contact angle}b. The particles were pre-dispersed in the bulk liquid and migrated towards the liquid-liquid interface, which means an advancing contact angle was measured as this best mimics the sample preparation conditions. For the smooth particles, a contact angle of approximately 50$\degree$ was obtained. All three types of rough particles migrated completely into the secondary liquid, indicating a contact angle close to 0$\degree$. The increased hydrophilicity of the rough particles is in accordance with  Wenzel wetting (Equation \ref{eq: Yan contact angle}). The roughness factor $r_w$ for each particle type was calculated from the AFM profiles in Figure \ref{fig: Surface profiles and contact angle}, with all values being close to the theoretical value of 1.9 for a sphere covered by closely-packed hemispherical asperities.~\cite{Yan2007} According to Equation \ref{eq: Yan contact angle}, a roughness factor of 1.9 would result in a value of 0$\degree$ for the rough particle given a smooth particle contact angle at or below 58$\degree$. There are, however, a few caveats to this contact angle measurement. First, the contact angle of a capillary bridge in the capillary suspension can be different than the contact angle observed at the liquid-liquid interface. This difference was already observed by Bossler et al.,~\cite{Bossler2016} with a lower contact angle in the capillary bridges compared to the angle at the flat interface due to the low porosity of the particles. However, the resolution of the microscope is not high enough to extract the contact angle directly from the liquid bridges between 3 $\mu$m particles, which is why the liquid-liquid interface method was chosen. Second, Equation \ref{eq: Yan contact angle} is only valid for Wenzel wetting behavior in which the asperities are fully wetted. As will be discussed in the next section, the secondary liquid first fills the asperities before forming capillary bridges, making it more likely that the assumption of a Wenzel wetting state is valid for these rough particles. A presumably lower contact angle would lead to a higher capillary force for the rough particles, but the Wenzel equation neglects the local surface roughness and predicts an equlibrium angle which is not necessary present in the bridge. Finally, particle roughness can lead to contact line pinning, leaving particles at metastable states at the liquid-liquid interface. A calculation of the free energy increase needed to depin the liquid-liquid contact line from a single asperity, as was done in Zanini et al.~\cite{Zanini2017}, would indeed suggest that pinning is possible, as this free energy is a factor $10^4 \sim 10^6$, depending on the asperity height, higher than $k_B T$.  However, pinning was only observed for a very low number of rough particles of type R100 and R250, with the vast majority being fully engulfed by the secondary liquid.

\subsection{Linking microstructure and rheological response}
\subsubsection{Low secondary liquid volume} 
Without secondary liquid, weak networks due to van der Waals interactions were obtained for all particle type suspensions. As a consequence, no settling was observed. The plateau storage modulus of the suspension with smooth particles was slightly larger compared to the three rough ones, with a value around $7 \times 10^2$ Pa compared to $3 \times 10^2$ Pa for R250, as shown in supplemental Figure S4. The values are too close to the low-torque limit of the rheometer for a more accurate estimation. To describe the microstructure of each sample, semilocal measures of the coordination number $z$ and clustering coefficient $c$ are detected from the confocal images. The coordination number corresponds to the number of bridges per particle, but is not sufficient by itself to describe the complex structure of the particle network at the low particle volume fraction used in this work. Thus, this measure is supplemented by the clustering coefficient. The average coordination numbers and clustering coefficients are summarized in Tables \ref{tbl:coordnr saos} and \ref{tbl:cluster saos}. It can be seen that the rougher particles tend to form less clustered networks, indicated by the lower coordination numbers and clustering coefficients. This trend is consistent with earlier research on suspensions and colloidal gels.~\cite{Davis2003,Lin2021,Nguyen2020sim} When two particles with a rough surface approach each other, the asperities can hinder the lubricative interactions leading to a weaker connection between the particles.~\cite{Davis2003,Lin2021} Moreover, the presence of rolling friction might limit the rearrangement of the particles into denser clusters, leading to more string-like networks with lower coordination numbers.~\cite{Nguyen2020sim} 
\begin{table}[tbp]
\small
  \caption{\ Average coordination numbers z for capillary suspensions with varying secondary liquid contents and varying particle roughness}
  \label{tbl:coordnr saos}
  \begin{tabular}{lllll}
    \toprule
    $\phi_\mathrm{{sec}}$ & $z_\mathrm{{Smooth}}$ & $z_\mathrm{{R40}}$ & $z_\mathrm{{R100}}$ & $z_\mathrm{{R250}}$\\
    
    [vol\%] & [-] & [-] & [-] & [-] \\
    \midrule
    0 & 3.7 $\pm$ 0.3 & 4.0 $\pm$ 0.3 & 3.5 $\pm$ 0.1 & 3.5 $\pm$ 0.1\\
    0.1 & 4.9 $\pm$ 0.1 & 4.6 $\pm$ 0.1 & 4.0 $\pm$ 0.1 & 3.3 $\pm$ 0.1\\
    0.2 & 4.9 $\pm$ 0.1 & 4.6 $\pm$ 0.3 & 3.9 $\pm$ 0.1 & 3.3 $\pm$ 0.1\\
    1.0 & 5.4 $\pm$ 0.1 & 5.2 $\pm$ 0.4 & 4.1 $\pm$ 0.1 & 3.2 $\pm$ 0.2\\
    3.0 & - & - & 5.0 $\pm$ 0.1 & -\\
    5.0 & - & - & - & 3.7 $\pm$ 0.3\\ 
    \bottomrule
  \end{tabular}
\end{table}

\begin{figure}[tbp]
    \centering
    \includegraphics[width=0.75\textwidth]{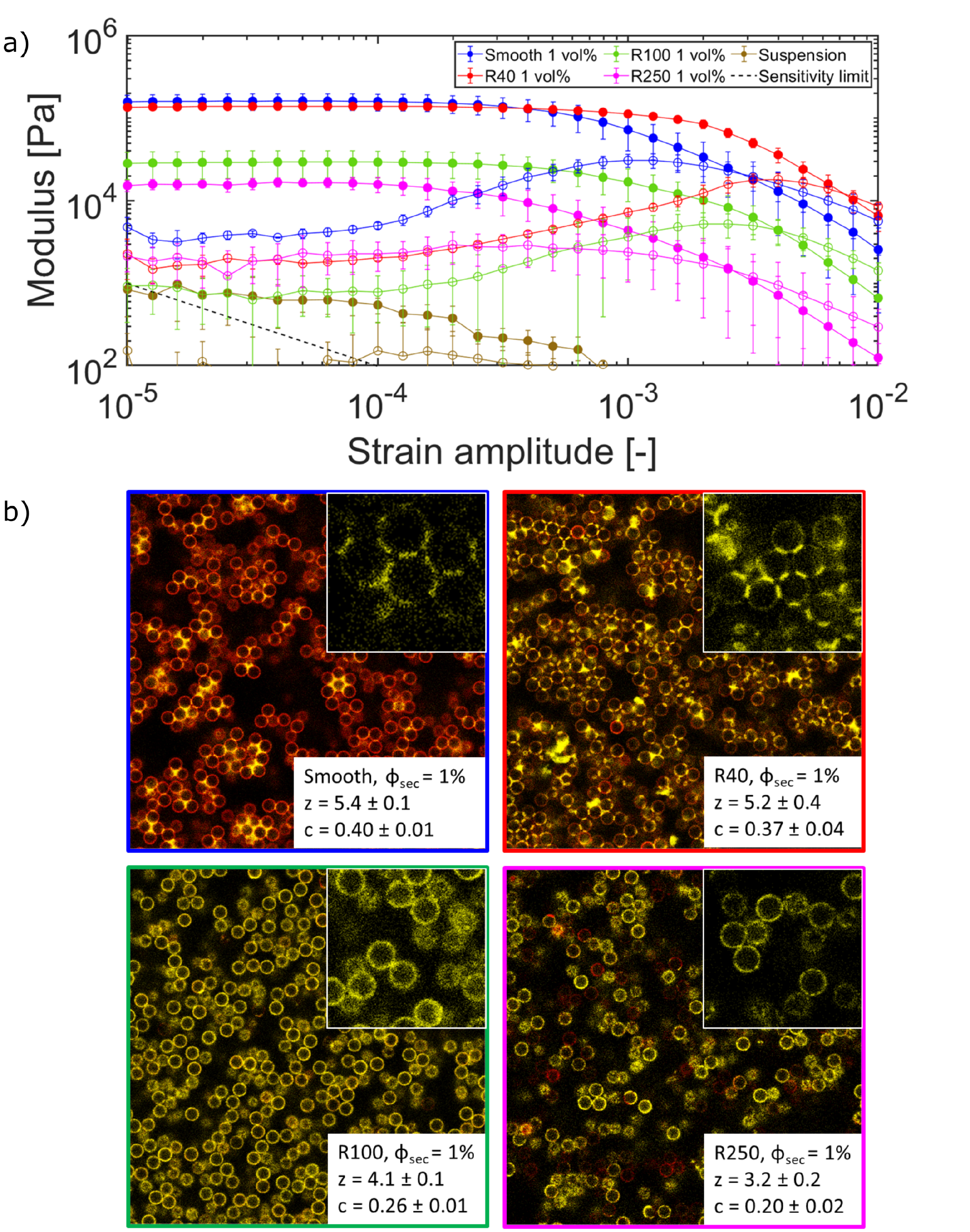}
    \caption{a) Amplitude sweeps and b) corresponding confocal micrographs of capillary suspensions with particles of different roughness and $\phi_\mathrm{{solid}} = 20 ~\mathrm{vol\%}$, $\phi_\mathrm{{sec}}= 1  ~\mathrm{vol\%}$. Filled symbols: $G'$, open symbols: $G''$. The dotted line shows the sensitivity limit calculated from the low torque limit provided by the manufacturer. Confocal micrographs in b) have a dimension of 82 $\mu$m x 82 $\mu$m. The particles are shown as red rings and the secondary fluid is colored yellow. Insets show a magnified image of representative bridges.}
    \label{fig:Roughness rheology same phisec}
\end{figure}

\begin{table}[tbp]
\small
  \caption{\ Average clustering coefficients c for capillary suspensions with varying secondary liquid contents and varying particle roughness}
  \label{tbl:cluster saos}
  \begin{tabular}{lllll}
    \toprule
    $\phi_\mathrm{{sec}}$ & $c_\mathrm{{Smooth}}$ & $c_\mathrm{{R40}}$ & $c_\mathrm{{R100}}$ & $c_\mathrm{{R250}}$\\
    
    [vol\%] & [-] & [-] & [-] & [-] \\
    \midrule
    0 & 0.27 $\pm$ 0.01 & 0.27 $\pm$ 0.01 & 0.21 $\pm$ 0.01 & 0.22 $\pm$ 0.01\\
    0.1 & 0.33 $\pm$ 0.01 & 0.31 $\pm$ 0.01 & 0.25 $\pm$ 0.01 & 0.26 $\pm$ 0.01\\
    0.2 & 0.34 $\pm$ 0.01 & 0.28 $\pm$ 0.02 & 0.25 $\pm$ 0.01 & 0.20 $\pm$ 0.01\\
    1.0 & 0.40 $\pm$ 0.01 & 0.37 $\pm$ 0.04 & 0.26 $\pm$ 0.01 & 0.20 $\pm$ 0.02\\
    3.0 & - & - & 0.30 $\pm$ 0.01 & -\\
    5.0 & - & - & - & 0.25 $\pm$ 0.04\\ 
    \bottomrule
  \end{tabular}
\end{table}
Adding $\phi_\mathrm{{sec}} = 1 ~\mathrm{vol\%}$ of secondary liquid significantly increases the viscoelastic moduli regardless of the particle roughness, as shown in Figure \ref{fig:Roughness rheology same phisec}. 
Since the smooth and rough particles are hydrophilic and with the aqueous phase as the bridging liquid, pendular state capillary suspensions with binary bridges are expected. The amplitude sweeps and corresponding microscope slices for the $\phi_\mathrm{{sec}}=1 ~\mathrm{vol\%}$ samples are shown in Figure \ref{fig:Roughness rheology same phisec}. All capillary suspensions show an increase of 1 or 2 decades in plateau storage modulus compared to the binary suspension. The coordination number and clustering coefficient also increases relative to the $\phi_\mathrm{{sec}}=0 ~\mathrm{vol\%}$ case for all samples except R250, which shows that the secondary liquid plays an active role in determining the network structure rather than just reinforcing the van der Waals connections. Both $z$ and $c$ decrease with increasing roughness. For different amounts of secondary fluid between $\phi_\mathrm{{sec}}= 0.1 ~\mathrm{vol\%}$ and $1 ~\mathrm{vol\%}$, a similar trend is visible in Tables \ref{tbl:coordnr saos} and \ref{tbl:cluster saos}. 

The influence of adding 1 vol\% secondary fluid, however, is more pronounced for smoother particles than for rough particles, as seen in Figure \ref{fig:Roughness rheology same phisec}a. The high roughness samples, R100 and R250, have significantly lower moduli compared to the smooth and R40 particles implying a shift from the spherical regime toward the roughness or asperity wetting regimes with increasing roughness. The R40 capillary suspension has a nearly identical plateau storage modulus ($G_0 = 1.4 \times 10^{5}$ Pa) as the smooth particle capillary suspension ($G_0 = 1.6 \times 10^{5}$ Pa), but longer linear viscoelastic (LVE) region, as determined via the 5 \% deviation rule. The R250 sample has a shorter LVE region ($\gamma_\mathrm{crit} = 1.3 \times 10^{-4}$) compared to the smooth ($\gamma_\mathrm{crit} = 1.8 \times 10^{-4}$). Interestingly, the R100 sample already has a slightly longer LVE region ($\gamma_\mathrm{crit} = 2.6 \times 10^{-4}$) than the smooth particle capillary suspension, even though the storage modulus decreases, which suggests that the roughness influences the critical strain. The decrease in $G'$ is accompanied by a simultaneous increase in $G''$, which shows a maximum according to type III LAOS behavior following the classification of Hyun et al.~\cite{Hyun2011} The specific reason for this $G''$ maximum in capillary suspensions is still unknown, but the position of the peak follows the same trend as the critical strain (i.e. R250 $<$ smooth $<$ R100 $<$ R40).  

The reason for these rheological changes can be explained by the confocal micrographs (Figure~\ref{fig:Roughness rheology same phisec}b). A close-up of the liquid bridges is shown in the insets in the top right corner of each of the confocal micrographs in Figure \ref{fig:Roughness rheology same phisec}b. In the low roughness samples, the pendular bridges (yellow) can be seen clearly connecting individual particles (red rings). For the high roughness samples, a liquid film surrounds the particles without forming a visible bridge. The reason for this change is that the secondary liquid volume is not sufficient to completely fill the asperities, preventing the formation of a single pendular bridge (spherical regime). This situation, where only small capillary bridges between the asperities are formed (asperity or roughness regime), results in a weaker capillary force than for a pendular bridge. The lower interparticle force explains why the viscoelastic moduli of the high roughness suspensions are lower relative to the smooth particles. This is also visible in the resulting structure where the coordination number and clustering coefficient decrease as the roughness increases. However, contrary to the other particle systems, there is no clear trend with increasing $\phi_\mathrm{{sec}}$ visible for the rough R250 particles. Hence, the secondary liquid influences both the number and strength of the liquid bridges, which in turn affects the rheological properties. More specifically, we expect that the capillary force of the bridges and the network structure determine the magnitude of the viscoelastic moduli, while the combination of the bridge strength and the particle contact, i.e. friction, determines the critical strain, at which point some relative particle movement occurs in the network. To test this hypothesis, the secondary fluid volume can be increased for the rough particles to ensure that all systems are in the spherical regime.

\subsubsection{Adjusted secondary liquid volume}
\begin{figure}[tbp]
    \centering
    \includegraphics[width=0.75\textwidth]{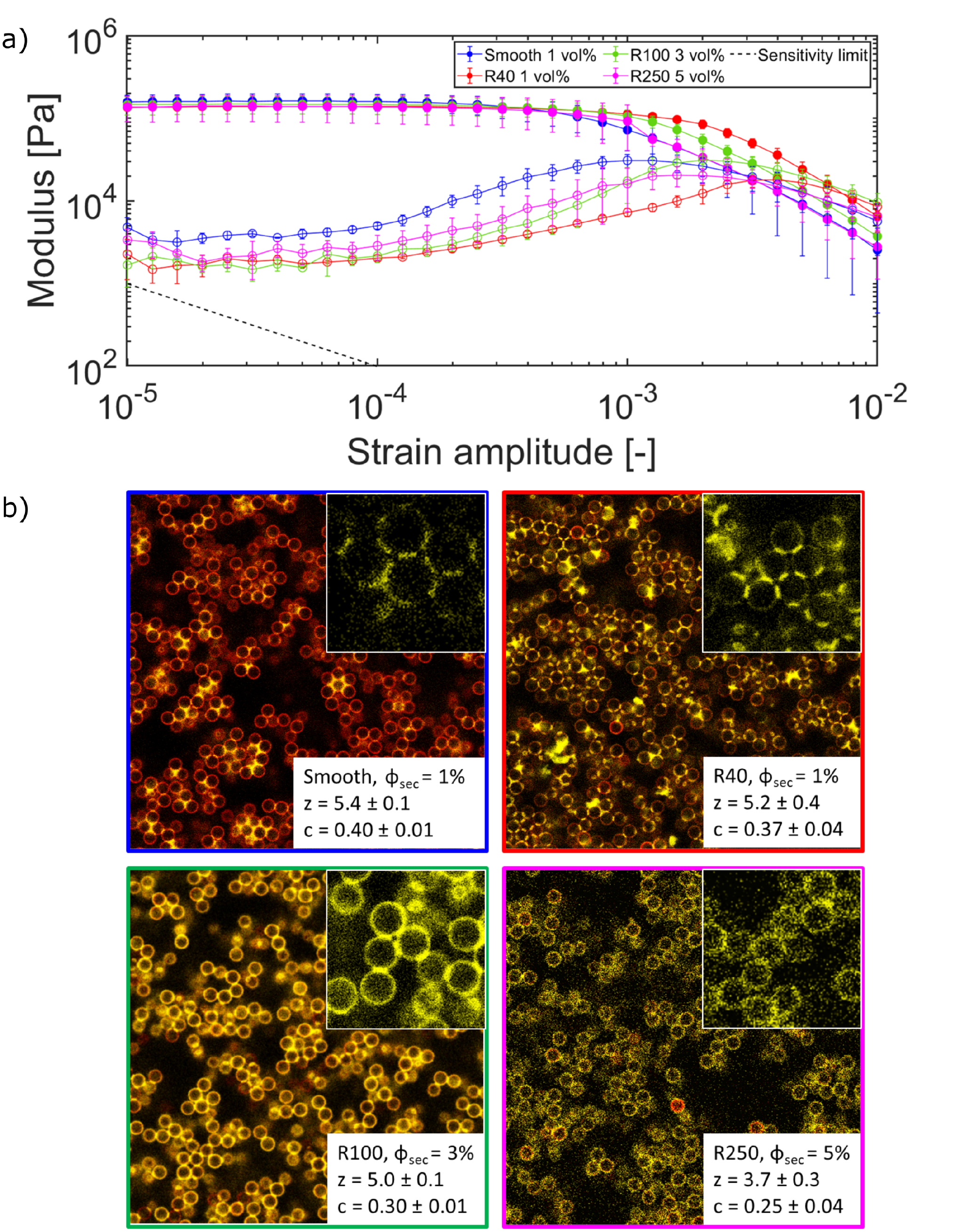}
    \caption{a) Amplitude sweeps and b) corresponding confocal micrographs of capillary suspensions with particles of different roughness and $\phi_\mathrm{{solid}} = 20 ~\mathrm{vol\%}$, adjusted $\phi_\mathrm{{sec}}$. Filled symbols in a) represent $G'$ and open symbols $G''$. The dotted line shows the sensitivity limit calculated from the low torque limit provided by the manufacturer. Confocal micrographs in b) have a dimension of 82 $\mu$m $\times$ 82 $\mu$m. The particles are shown as red rings and the secondary fluid is colored yellow. Insets show a magnified image of representative bridges.}
    \label{fig:Roughness rheology adj phisec}
\end{figure}

By adjusting the secondary fluid volume for the high roughness samples, the storage modulus in all samples can be matched, as seen on Figure~\ref{fig:Roughness rheology adj phisec}a.
To do this, we simply added the extra volume needed to fill the asperities, which resulted in a composition with $\phi_\mathrm{{sec}}=3 ~\mathrm{vol\%}$ for R100 and $\phi_\mathrm{{sec}}=5 ~\mathrm{vol\%}$ for R250. After the addition of the extra liquid, these two samples also show toroidal bridges on the confocal micrographs (Figure~\ref{fig:Roughness rheology adj phisec}b). There remains, however, an increase in the critical strain, i.e. the end of the LVE region, for all rough samples compared to the smooth sample. The peak in the loss modulus also appears at lower strain for the smooth particles. The increased yield strain for all three rough samples ($\gamma_\mathrm{crit} = 2.3 \times 10^{-4}$ to $3.7 \times 10^{-4}$) compared to the smooth particle system ($\gamma_\mathrm{crit} = 1.8 \times 10^{-4}$) must therefore be due to the physical presence of the roughness. If the yield stress is evaluated as the stress at the end of the LVE region, the rough particles have a higher yield stress ($\sigma_{y} \approx 41-54$ Pa) than the smooth particle system ($\sigma_{y} \approx 31$ Pa) due to their higher yield strain. The increased strength can be caused by an additional frictional contribution due to the roughness (that is a tangential contact force due to interlocking asperities) on top of the capillary force that has to be overcome before particle-particle movement can occur. Besides a change in sliding friction, the presence of particle asperities can introduce a strong increase in rolling friction as well, which has been shown to enhance the rigid body movement of particle clusters and reduce relative particle movement.~\cite{Hsu2018,Singh2020} Alternatively, the increased contact angle hysteresis can be causing a higher capillary torque of the bridges.~\cite{Naga2021} Frictional dissipation is expected to show up in the loss modulus, which is lower for all three rough particles compared to the smooth particles, possibly caused by interlocking structures or a lower total contribution of friction due to the lower values of $z$ and $c$. The crossover strain is approximately equal for smooth, R100 and R250 particles ($\gamma_\mathrm{cross} \approx 3.0 \times 10^{-3}$). Because the particle size and bridge volume are fairly similar due to the adjusted $\phi_\mathrm{{sec}}$, it might be logical that the fluidization point is the same after rupturing enough bridges. However, in order to conclude this with certainty, simultaneous imaging and rheology is required. 

The increased resistance to particle movement due to the roughness is qualitatively confirmed by preliminary experiments with a rheometer mounted to a fast scanning confocal microscope, which is shown in supplemental Figure S5 and supplemental videos 1 to 3. The smooth particle capillary suspension shows more relative particle movement around the $G''$ peak compared to the rough ones, whereas the rough particles exhibit a more pronounced rigid body movement until the crossover point occurs. Furthermore, the mobility of the liquid phase after a bridge breaks is higher for the smooth particles, that is the liquid is transported over the particle surface and reforms a new bridge on a neighboring particle. We show two examples in supplemental videos 4 and 5.  

\begin{figure}[tbp]
    \centering
    \includegraphics[width=0.75\textwidth]{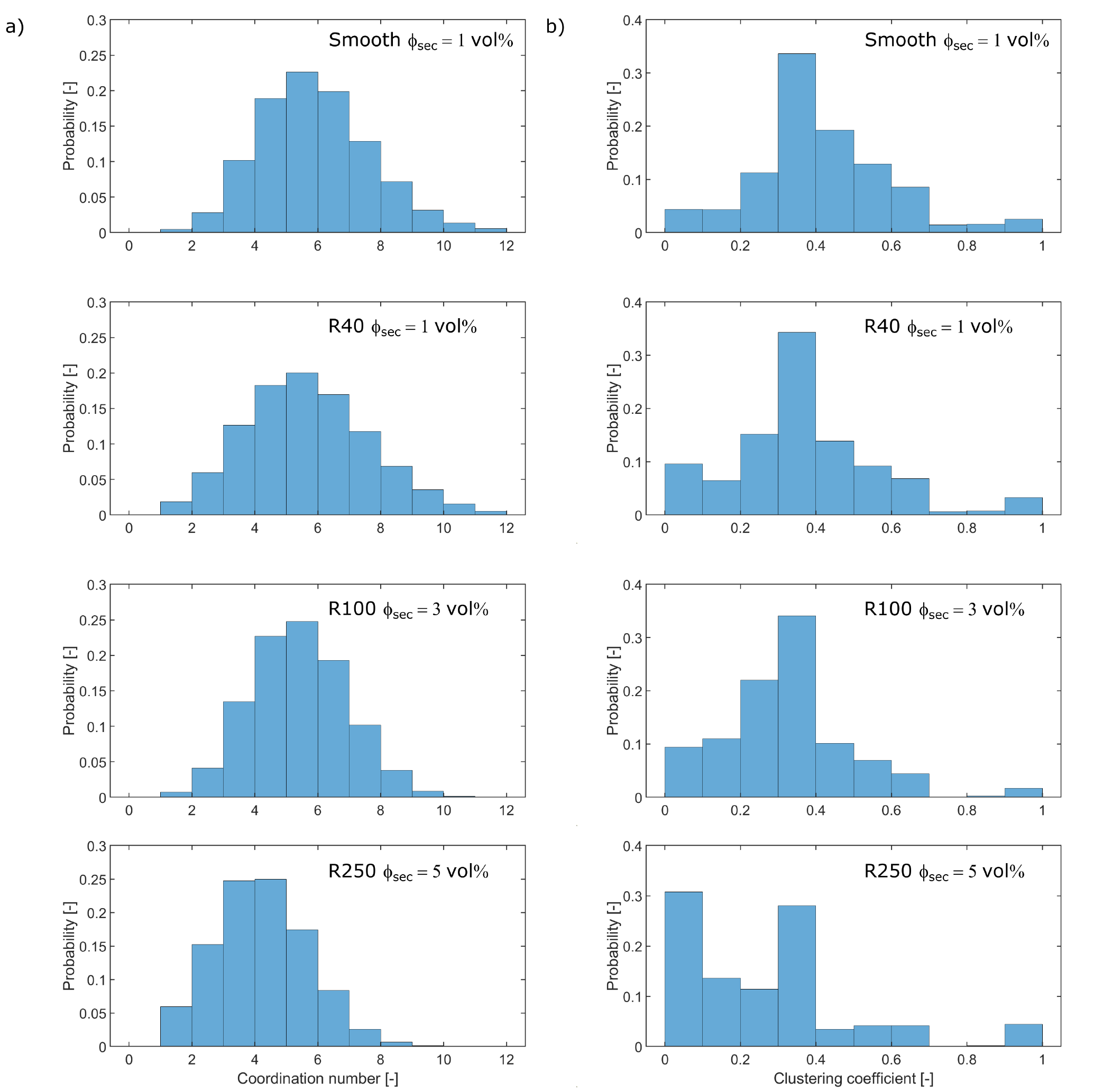}
    \caption{a) Coordination number and b) clustering coefficient histograms for the capillary suspensions with particles of different roughness and $\phi_\mathrm{{solid}}= 20 ~\mathrm{vol\%}$, adjusted $\phi_\mathrm{{sec}}$.}
    \label{fig:Histograms}
\end{figure}

The trend of a decrease in the average coordination number and clustering coefficient with increasing roughness also prevails for the adjusted R100 and R250 samples. This is also visible in the probability distributions, shown on Figure \ref{fig:Histograms}. These histograms show the distributions behind the mean coordination numbers and clustering coefficients reported in Figure~\ref{fig:Roughness rheology adj phisec} and Tables \ref{tbl:coordnr saos} and \ref{tbl:cluster saos}. 
With increased particle roughness, the entire coordination number distribution shifts to lower values and the high coordination number tail disappears. For the R250 sample, a large increase in $z = 1$ and $z = 2$ can be seen. For the clustering coefficient, there is a gradual increase in $c < 0.2$ with a simultaneous decrease in $c > 0.4$. For the R250 sample, we see a sharp increase of $c = 0$ partially caused by the shift towards linear strings or small branches with one or two neighbors.

The absence of high coordination number and clustering coefficient values implies that no large particle aggregates are formed during the suspension preparation phase when a large secondary liquid droplet is present. For the rough particles, more secondary liquid is used to fill the asperities, which decreases the chance of a local excess of secondary liquid during the mixing process. This is qualitatively confirmed by confocal micrographs with a larger field of view (246 $\mu$m x 246 $\mu$m), shown in the supplemental images Figure S6 and Figure S7. In the large-scale micrographs of the smooth particle network, a few dense yellow clusters can be seen in addition to some larger regions without any particles, i.e. voids in the structure. The R250 capillary suspension, by contrast, shows a more homogeneous structure without larger aggregates or voids. While it is known that a secondary liquid volume corresponding to the funicular state denotes the maximum in the overall capillary suspension strength, and that increasing the amount of secondary liquid beyond this point weakens the sample by creating capillary aggregates~\cite{Bindgen2021}, the relative contribution of different types of bridges within a single network -- binary bridges, trimers or small aggregates -- to the sample strength has not yet been discovered and could be a very interesting topic for future research. It could be that the more homogeneous structure, for example of the R250 sample, also increases the viscoelastic moduli by avoiding any weak spots in the network and in this way compensates for the lower values of coordination number and clustering coefficient.~\cite{Johnson2019} Furthermore, increasing the particle roughness to obtain a more a homogeneous structure might be particularly interesting for applications in sintered ceramics~\cite{Dittmann2013,Dittmann2014,Dittmann2016}  where a uniform pore size can be desired, as well as in cases where the secondary liquid is hard to disperse. 

A side effect of the employed roughness procedure is an increase in effective particle size due to the adsorption of the nanoparticles. This size increase was also recognized by the particle detection algorithm, as the average detected radius increased from 1.5 $\mu$m to 1.7 $\mu$m. An increase in particle size decreases the E\"{o}tv\"{o}s number $\mathrm{Eo}$, also known as Bond number,
\begin{equation}
    \mathrm{Eo} = \frac{\Delta\rho g R^2}{2 \Gamma}
    \label{Eq: eotvos}
\end{equation}
where $\Delta\rho$ is the density difference between bulk and secondary liquid, $g$ the gravitational acceleration, $R$ the particle radius and $\Gamma$ the interfacial tension. The increasing particle size reduces the relative magnitude of the capillary force from $\mathrm{Eo}_\mathrm{smooth} = 5.9 \times 10^{-7}$ to $\mathrm{Eo}_\mathrm{R250} = 8.3 \times 10^{-7}$. Typically, capillary suspensions with larger particles have a lower yield stress ($\sigma_{y} \propto R^{-1}$) and can have a more granular packing.~\cite{Bossler2018} However, the larger, rough particles show the opposite trends. As determined in Figure~\ref{fig:Roughness rheology adj phisec}, the yield stress of the rough particles is higher than that of the smooth particles for matching relative bridge volumes. The coordination numbers for the rough particles are also lower compared to the smaller, smooth particles, which represents a less granular-like packing. In order to definitively conclude that the roughness effects dominate over the particle size increase, we prepared a capillary suspension with the same composition with smooth 10 $\mu$m silica particles ($\mathrm{Eo}_\mathrm{smooth} = 7.1 \times 10^{-6}$), bought from the same supplier. While a denser network was not observed ($z = 5.2$ and $c = 0.40$, very similar values to the smooth 3 $\mu$m sample), the yield stress, evaluated at the yield strain from an oscillatory test, followed the expected $1/R$ scaling for capillary suspensions. 
We can, thus, conclude from the clear differences in network structure and yield strain between rough and smooth particles that the particle roughness itself, and not just the size increase, causes changes in the properties of capillary suspensions. Since there is a clear difference in the structure of capillary suspensions prepared with rough particles, even when the plateau modulus and bridge size are matched, there must also be a change to the particle interactions.

\subsection{Medium amplitude oscillatory shear to investigate particle contacts}

To capture changes in the particle interactions, the influence of particle roughness was examined using information obtained from the medium amplitude (or asymptotically nonlinear) oscillatory shear regime (MAOS). By definition, the MAOS regime is the nonlinear region where only the first and third harmonic stress signals appear. Previous research has shown that the scaling of the third harmonic is non-integer and non-cubic~\cite{Natalia2020} and that it is sensitive to particle collisions~\cite{Natalia2022}. Since the particle collisions are expected to be indirectly affected by the roughness through the strength of the bridges and directly affected via the frictional interactions, the third harmonic response should vary between the smooth and rough particles. The definitions of the elastic ($\sigma'_3$) and viscous ($\sigma''_3$) third harmonic stress coefficients are given by the following equations:
\begin{equation}
    \sigma'_3(\omega,\gamma_0) = -[e_3](\omega) \cdot \gamma_0^{m_\mathrm{{3,elastic}}} + \mathcal{O}(\gamma_0^{p_{5,elastic}})
    \label{Eq: 3rd harmonic elastic}
\end{equation}
\begin{equation}
    \sigma''_3(\omega,\gamma_0) = \omega[v_3](\omega) \cdot \gamma_0^{m_\mathrm{{3,viscous}}} + \mathcal{O}(\gamma_0^{p_{5,viscous}})
    \label{Eq: 3rd harmonic viscous}
\end{equation}
where $\gamma_0$ is the applied strain, $e_3[\omega]$ and $v_3[\omega]$ are the elastic and viscous intrinsic nonlinear material functions, both of which depend on the applied angular frequency $\omega$. The leading nonlinear power law exponents are denoted with $m_\mathrm{{3,elastic}}$ and $m_\mathrm{{3,viscous}}$, and $p_{5,elastic}$ and $p_{5,viscous}$ are higher order exponents. 

\begin{figure}[tbp!]
    \centering
    \includegraphics[width=0.75\textwidth]{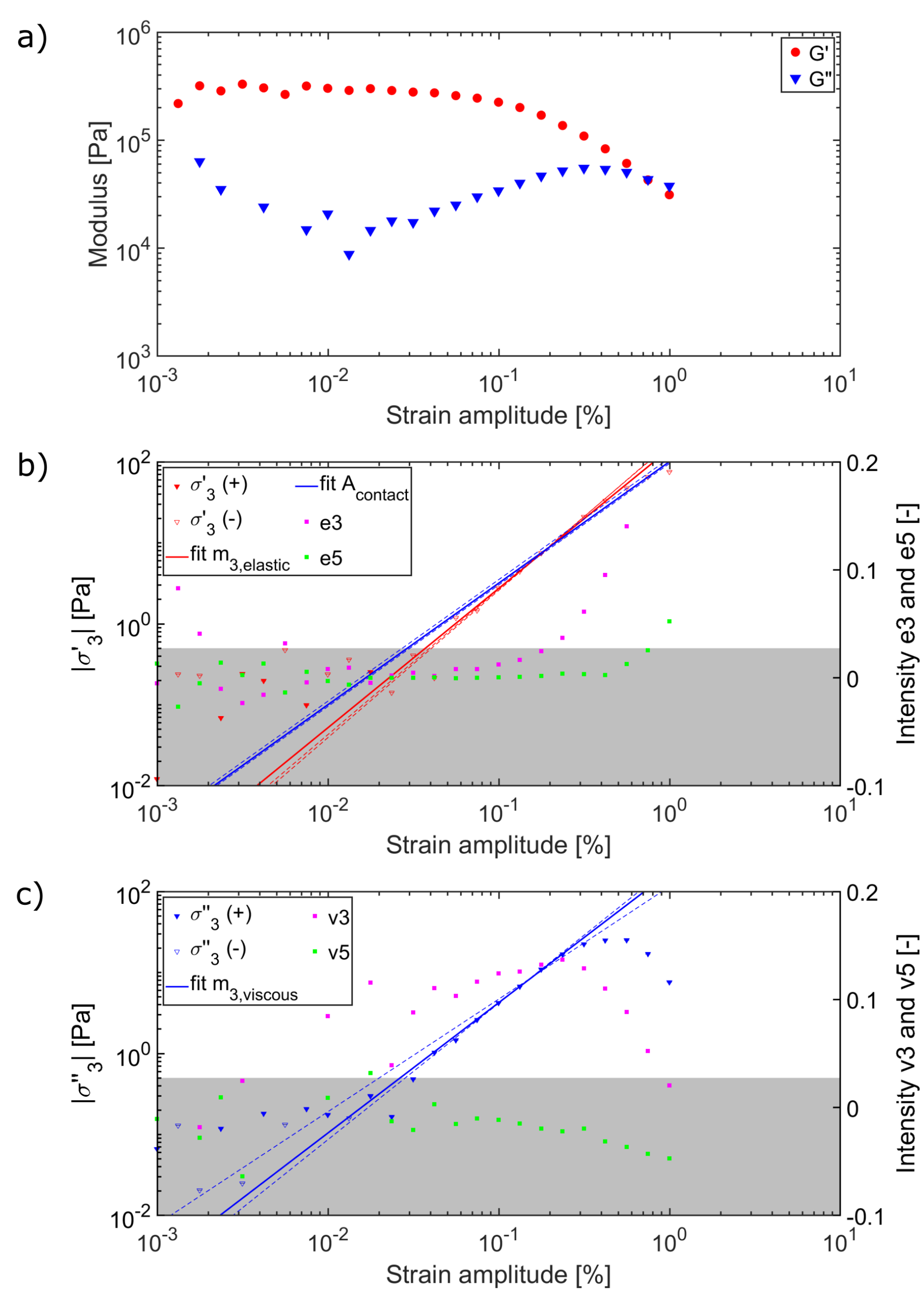}
    \caption{Example fit of the third harmonic elastic and viscous stress scalings in the asymptotically nonlinear regime for measurements on an R250 sample with an (a-c) 8 mm parallel plate geometry. The stress noise floor is shown as gray shading below 0.5 Pa. The dashed lines show the sensitivity of the fits by including or excluding an extra data point. The second Y-axis shows the relative magnitude of the third and fifth harmonic normalized by the first harmonic. A full version of this Figure including the comparison with a 25 mm parallel plate can be found in supplemental Figure S8}
    \label{fig:MAOS fit}
\end{figure}

One example fit of the power law scaling for the R250 sample with adjusted secondary liquid volume ($\phi_\mathrm{{sec}} = 5 ~\mathrm{vol\%}$) is shown in Figure \ref{fig:MAOS fit}. 
The amplitude sweep is shown in Figure \ref{fig:MAOS fit}a, and the third harmonics $\sigma'_3$ and $\sigma''_3$ in Figure \ref{fig:MAOS fit}b and Figure \ref{fig:MAOS fit}c, respectively, are plotted as a function of strain, together with the intensity of the third and fifth harmonic. The power law slopes $m_\mathrm{{3,elastic}}$ and $m_\mathrm{{3,viscous}}$ are determined according to Equations \ref{Eq: 3rd harmonic elastic} and \ref{Eq: 3rd harmonic viscous}. The fitting region is taken as the strain values between the noise floor and the maximum of the third harmonic viscous intensity. The magnitude of the fifth harmonic, $\sigma_5 / \sigma_1$, should be negligibly small in this region. As can be seen in the example shown in Figure \ref{fig:MAOS fit}, the viscous intensity $\sigma_5'' / \sigma_1''$ remains small while the elastic fifth harmonic $\sigma_5' / \sigma_1'$ only begins to rise near the end of the fitting region. Furthermore, at strains below the $G''$ maximum, the yielding of the network is reversible, as can be seen on supplemental Figure S9. The emergence of higher odd harmonics in this reversible yielding regime was also demonstrated in the works of Domenech and Velankar~\cite{Domenech2015} and Yang and Velankar~\cite{Yang2017}. Fitting this strain region results in values for $m_\mathrm{{3,elastic}}$ between 1.5 and 2.5. If the region in between the $G''$ maximum and the crossover point is included as well, this relaxes the scaling of $m_\mathrm{{3,elastic}}$ to values between 1.5 and 1, which is more similar to the pendular capillary suspension system in the work of Natalia et al.~\cite{Natalia2022}

Important to note is that the scaling that can be observed is limited by the sensitivity window of the rheometer, shown as gray shading in Figure \ref{fig:MAOS fit}. As a result of the smaller parallel plate geometry, the noise floor corresponding to the sensitivity limit of the rheometer is shifted to higher stress values, which could mask an initial cubic scaling, as already discussed by Natalia et al.~\cite{Natalia2022} For the interval shown in Figure \ref{fig:MAOS fit}, the elastic and viscous exponents are 1.7 $\pm$ 0.1 and 1.6 $\pm$ 0.2 using the 8 mm. The dashed lines show the effect on the fit after including or excluding an extra datapoint. Additional experiments conducted with a 25 mm parallel plate show very comparable results for $m_\mathrm{{3,elastic}}$ (1.7 $\pm$ 0.1) and $m_\mathrm{{3,viscous}}$ (1.8 $\pm$ 0.1), which are shown in supplemental Figure S9. The main difference between using the different-sized plates is that the LVE region is extended by one strain decade with the 25 mm plate, indicating that some slip is present for the 8 mm geometry. 

\begin{figure}[tbp]
    \centering
    \includegraphics[width=0.75\textwidth]{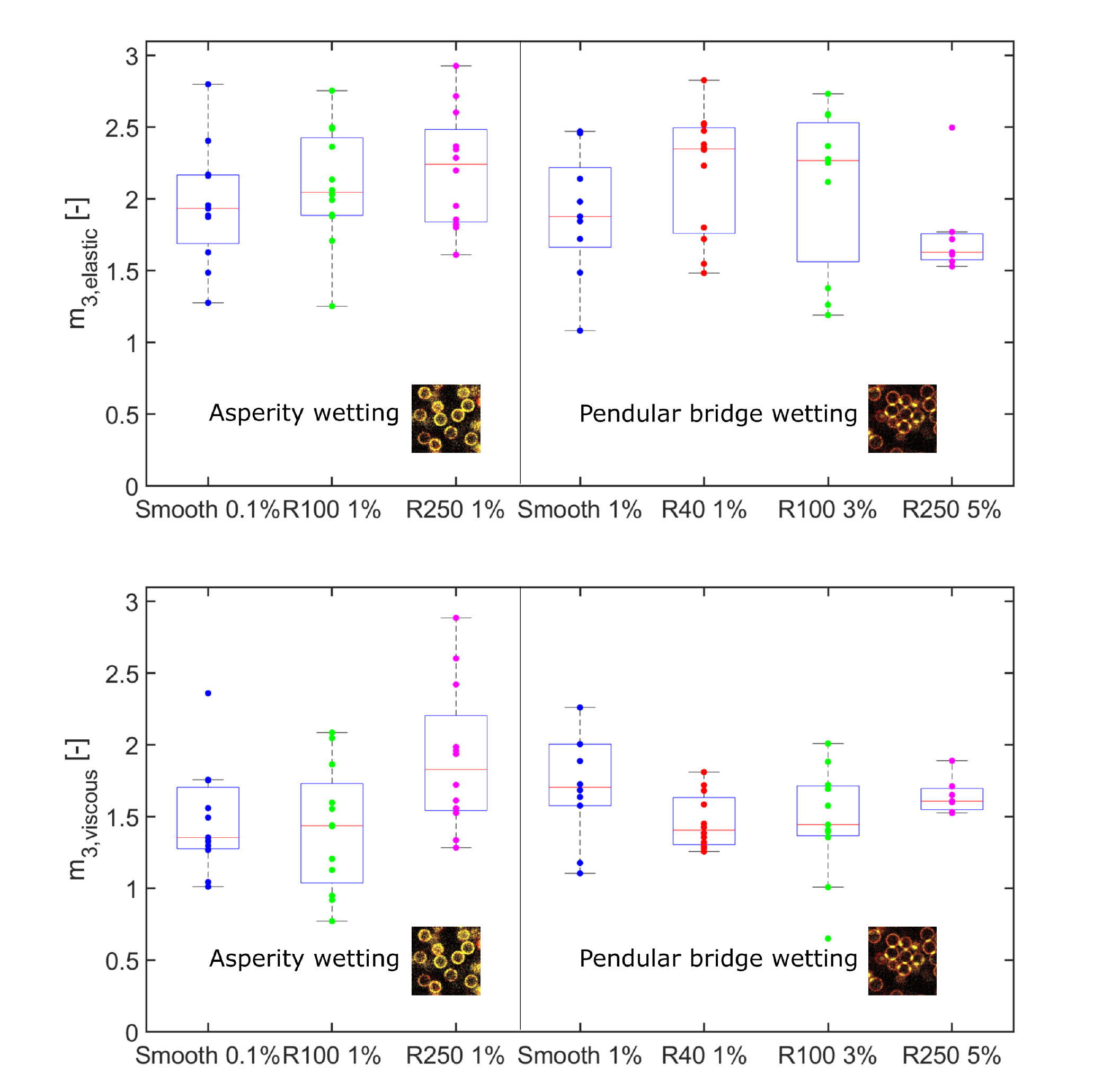}
    \caption{Boxplots with the distribution of $m_\mathrm{{3,elastic}}$ and $m_\mathrm{{3,viscous}}$ for capillary suspensions with different rough particles and secondary liquid volume. The box shows the interquartile range (IQR) 25\% - 75\% with mean line and error range from 1.5 $\times$ IQR.}
    \label{fig:Boxplot 8mm scalings}
\end{figure}

The results of the power law fitting of $\sigma'_3$ and $\sigma''_3$ for both the samples with $\phi_\mathrm{{sec}}=1 ~\mathrm{vol\%}$ and adjusted secondary fluid can be seen in Figure \ref{fig:Boxplot 8mm scalings}. 
First, both the scaling of $m_\mathrm{{3,elastic}}$ and $m_\mathrm{{3,viscous}}$ is noncubic and noninteger for all samples, in accordance with the works of Natalia et al.~\cite{Natalia2020,Natalia2022} The power law exponents do not differ significantly between sample groups, with $m_\mathrm{{3,elastic}}$ values falling mostly between 1.5 and 2.5 and $m_\mathrm{{3,viscous}}$ values between 1 and 2. The spread on the data is quite large due to the limited number of datapoints above the noise floor available for the fitting owing to the 8 mm parallel plate geometry. This is also the reason why the suspensions without secondary liquid were not measured: their non-linear stress response was below the detectable limit. 

A cubic scaling of $m_3$ is expected for both the elastic and viscous stresses for most materials, but a non-cubic, non-integer scaling is observed in capillary (and some other) suspensions instead due to the combination of the attractive capillary bridge (or van der Waals) force and the Hertzian repulsion. The resulting total force on a contacting particle pair (with identical properties) connected by a liquid bridge can then be described by~\cite{Butt2010}
\begin{equation}
F = -\frac{4}{3} E^{*} R^{*}{}^{1/2} \delta^{3/2} + 4 \pi \Gamma R^{*} \left(1 + \frac{\delta}{4r}\right)
    \label{Eq: capforce and Hertz}
\end{equation}
where $R^{*} = R/2 $ is the effective particle radius, $E^{*} = \frac{E}{2(1-\nu^2)} $ is the effective Young's modulus, $\delta$ is the indentation depth caused by the capillary force, $r$ is the principal radius of curvature for the bridge and $\Gamma$ is the interfacial tension. The negative term of Equation \ref{Eq: capforce and Hertz} denotes the Hertzian repulsion and the positive term is the attractive capillary force. The elastic scaling is presumed to be caused by this combination of a repulsive Hertzian-like particle contact induced by the attractive capillary force in combination with the applied shear deformation. Therefore, Natalia et al.\@ proposed the following equation to relate the third harmonic elastic stress $\sigma'_3$ to the contact strength $A$ and the precompression strain $\hat{\gamma}$: 
\begin{equation}
    \sigma'_3 =
    -A (\gamma_0 + \hat{\gamma})^{3/2} + A \hat{\gamma}^{3/2} \textrm{ for } ~\hat{\gamma} > 0 %\\
    \label{Eq: Acontact and gammahat}
\end{equation}
where $\gamma_0$ is the applied strain. In this equation, $A$ should be a measure for the Hertzian repulsion strength, whereas $\hat{\gamma}$ is the strain to separate particles already in contact. This equation predicts a power law exponent of 1.5 for the scaling of $\sigma'_3$ in the limit $\gamma_0 \gg \hat{\gamma}$. Equation \ref{Eq: Acontact and gammahat} is fitted to the $\sigma'_3$ data shown in Figure \ref{fig:MAOS fit} to obtain information about the particle contacts (blue lines). Again, the dashed lines show the sensitivity of the fitting method.

\begin{figure}[t]
    \centering
    \includegraphics[width=0.75\textwidth]{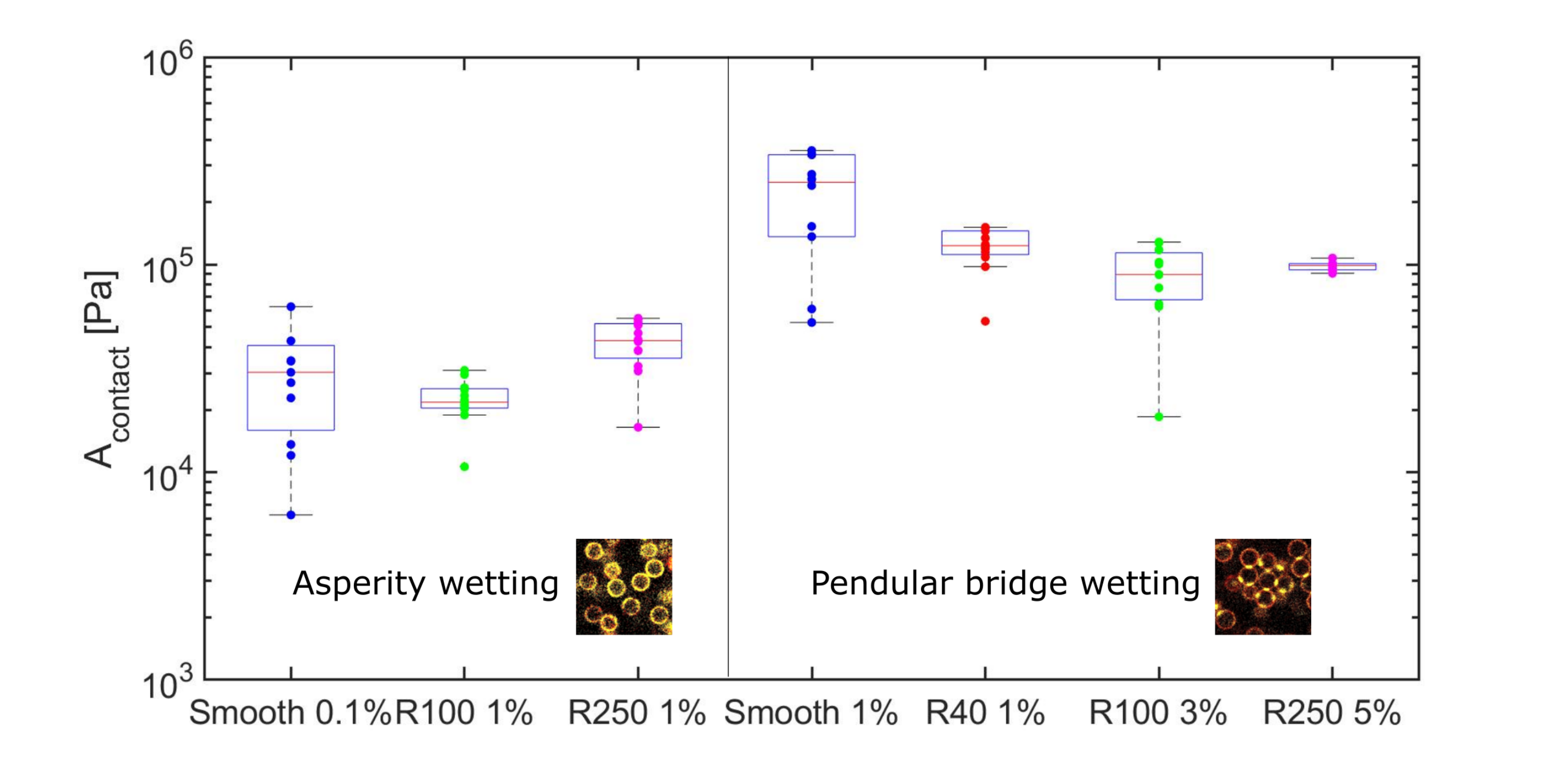}
    \caption{Boxplot of the magnitude of the repulsive Hertzian force $A$ for capillary suspensions with different rough particles and secondary liquid volume. The box shows the interquartile range (IQR) 25\% - 75\% with mean line and error range from 1.5 $\times$ IQR.}
    \label{fig:Boxplot 8mm Acontact}
\end{figure}

Interestingly, clear differences between the sample groups are obtained when the repulsive Hertzian contact parameter $A$ is plotted in Figure \ref{fig:Boxplot 8mm Acontact}. 
The magnitude of this repulsion is affected by the Young's modulus and the contact area, which in turn is affected by the magnitude of the capillary bridge force (Equation \ref{Eq: capforce and Hertz}). Concentrating on the matched $\phi_\mathrm{{sec}}$ samples, we can see that $A$ is largest for the smooth particle capillary suspension, and has a slightly lower value equal for the three rough particle samples. This trend is remarkable since their plateau storage moduli were equal within experimental accuracy, as shown on Figure \ref{fig:Roughness rheology adj phisec}. The possible explanation for this trend might be twofold. First, the rough particles are covered with a slightly porous St\"{o}ber layer with lower Young's modulus, whereas the purchased silica particles are uniformly hard. To test this hypothesis, we performed an additional measurement with smooth particles covered with a St\"{o}ber layer applied in the same manner as in the rough particle synthesis. The average and maximum $A$-value for the St\"{o}ber-covered smooth particles decreased slightly, but still remained higher than the rough particles. Second, the roughness can decrease the capillary force strength in addition to increasing the contact angle hysteresis.~\cite{Butt2009,Nguyen2021} This might cause a lower indentation depth for the rough particles. Unfortunately, the resolution of the microscope is not sufficient to properly extract the bridge shape profile and calculate the bridge strength. Using the ImageJ-software, we very roughly measured the bridge width and height and determined the relative bridge volume according to a concave arc of circumference with the assumption of zero particle separation.~\cite{Megias-Alguacil2009} With increasing roughness, the bridge volume, normalized by the particle size, decreases slightly from $V_\mathrm{{bridge}}/V_\mathrm{{part}}$ = 0.07 $\pm$ 0.01 for the smooth particles to $V_\mathrm{{bridge}}/V_\mathrm{{part}}$ = 0.05 $\pm$ 0.01 for R100 and R250 particles. This indicates the meniscus shape for smooth and rough particles is different, but drawing conclusions about the bridge strength purely from the volume is unwise since the contact angle in the bridge is unknown and likely different for each particle type. Furthermore, the fit values for $A_\mathrm{{contact}}$ are an average of all the bridges in the system and, as shown on Figure \ref{fig:Roughness rheology adj phisec}, the particles have fewer neighbors and a less clustered structure for the rough particles compared to the smooth particles. Compared to the matched $\phi_\mathrm{{sec}}$ samples with pendular bridges, the three samples with lower secondary liquid volume have significantly lower values of $A_{\mathrm{contact}}$. These samples exhibited asperity wetting since not enough liquid was available for bridge formation. The reduction in $A_{\mathrm{contact}}$ here is clearly caused by the lower bridge strength, which was also reflected in their lower viscoelastic moduli. It is important to note, however, that the parameter $A_{\mathrm{contact}}$, measured from the third harmonic in the asymptotically nonlinear regime, does not have a simple one-to-one correlation with the storage modulus from the first harmonic in the linear regime. For example, when the R250 and smooth samples were repeated with the roughened PP25, the moduli of the rough particle suspension were higher ($G'_{\mathrm{R250}} = 1.5 \times 10^{5}$ Pa versus $G'_{\mathrm{Smooth}} = 8.8 \times 10^4$ Pa) , but the strength of the Hertzian repulsion was still lower ($A_{\mathrm{Smooth}} = 2.4 \times 10^4$ Pa versus $A_{\mathrm{R250}} = 1.3 \times 10^4$ Pa), indicating that this parameter is indeed sensitive to the particle contacts. The precompression strain $\hat{\gamma}$ is more sensitive to the fitting procedure and all obtained values were scattered around zero. Therefore, they are not shown here and were not used to analyze the data.  

The viscous scaling should be related to dissipative frictional contacts in the system. For low normal forces and few contacting asperities, Hertzian contact theory predicts an elastic deformation of the asperities leading to a friction force $F_f$ that is proportional to the normal force $F_N^{2/3}$, which is also known as adhesion-controlled friction. In case of a high normal force or many asperities, Amonton's law applies leading to load-controlled friction with $F_f \propto F_N$. For capillary suspensions, the normal force is given by the capillary bridge force combined with the influence of the applied deformation.~\cite{Natalia2022} This leads to a relationship between the elastic and viscous third harmonic power law scalings with either
\begin{equation}
    m_\mathrm{{3,viscous}} \propto m_\mathrm{{3,elastic}}
    \label{Eq: Load control}
\end{equation}
for load-controlled friction, or
\begin{equation}
    m_\mathrm{{3,viscous}} \propto m_\mathrm{{3,elastic}}^{2/3}
    \label{Eq: Adhesion control}
\end{equation}
for adhesion-controlled friction. The relative scaling of smooth particles in both the pendular and capillary state in the work of Natalia et al.\@ tended towards Equation ~\ref{Eq: Adhesion control}.

\begin{figure}[tbp]
    \centering
    \includegraphics[width=0.75\textwidth]{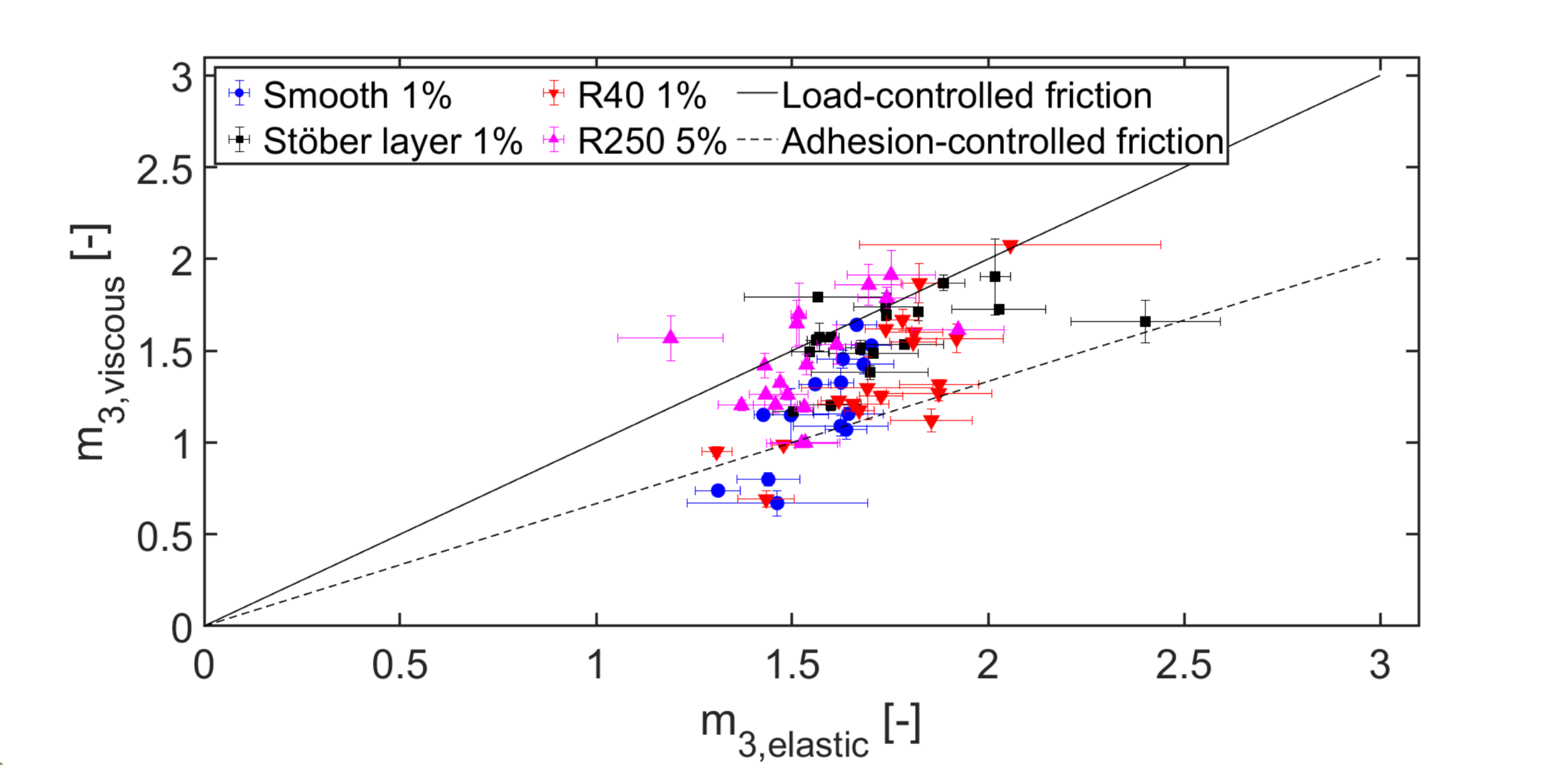}
    \caption{Relative scaling of $m_\mathrm{{3,elastic}}$ and $m_\mathrm{{3,viscous}}$ for capillary suspensions with different rough particles and secondary liquid volume. Measurements are performed with sandpaper-covered PP25 geometry. }
    \label{fig: Relative scaling PP25s}
\end{figure}

The relative scaling of the power law exponents for the repeat experiments with the sandpaper-covered PP25 geometry is shown in Figure \ref{fig: Relative scaling PP25s}. 
The diagonal lines show the expected load-controlled and adhesion-controlled scaling according to Equations \ref{Eq: Load control} and \ref{Eq: Adhesion control}. No sample exhibits purely load- or adhesion-controlled behavior and most points are situated in between the two diagonals, even for the smooth particles. The switch to St\"{o}ber covered smooth particles resulted in a slight increase in $m_\mathrm{{3,viscous}}$/$m_\mathrm{{3,elastic}}$ towards unity. This is in contrast to the results of Natalia et al.,~\cite{Natalia2022} where the adhesion-controlled friction dominated. However, a Student's t-test with a 95 \% confidence interval shows that the smooth particle data could come from the adhesion-controlled scaling (p-value = 0.052), whereas the R250 sample could come from the load-controlled scaling (p-value = 0.105), indicating the expected switch to load-controlled friction for rough particles where multiple asperities are in contact. The data from the other two samples did not fall in the 95 \% confidence interval of either of the two scalings. While these results do indicate a likely change from adhesion-controlled friction to load-controlled friction, additional measurements with larger sample sizes (increased signal to noise ratio) would be required to confirm this conclusion.

\section{Conclusions}
Using a combination of oscillatory rheology and confocal microscopy measurements, we have shown that particle roughness influences the capillary force in capillary suspensions leading to changes in structure. With increased roughness, a higher wetting liquid volume is required to fill the asperities on the particle surface. The capillary force increases when transitioning from asperity wetting to bridge wetting. As a consequence, the viscoelastic moduli of capillary suspensions with toroidal bridges are larger and the LVE region is extended compared to asperity wetting samples. By increasing the secondary liquid volume for high roughness samples, the viscoelastic moduli can be matched to that of the smooth particles, but the critical strain is slightly higher in all rough particle measurements. Interestingly, these almost identical rheological properties are obtained with a less clustered, more homogeneous microstructure for increased roughness samples, which shows that the particle roughness can be used as an extra tool to tune the structural and rheological properties of capillary suspensions. 

The influence of particle roughness was further examined using information obtained from the medium amplitude (or asymptotically nonlinear) oscillatory shear regime. The non-cubic and non-integer scaling of the third harmonic stress signals, originally found by Natalia et al., was confirmed here for smooth and rough particle capillary suspensions.~\cite{Natalia2020} For the third harmonic elastic stress scaling, we mostly obtain values between 1.5 and 2, while the third harmonic viscous stress is situated between 1 and 2 for both smooth and rough particles. The elastic scaling is presumed to be caused by the combination of a repulsive Hertzian-like particle contact induced by the attractive capillary force and the applied shear deformation. The strength of the Hertzian contact decreases with roughness when the samples transition from bridge to asperity wetting but increases to similar values for the samples where the secondary fluid volume is adjusted to provide for pendular bridges. This demonstrates the sensitivity of this rheological measure to the particle contact strength. The viscous scaling should be related to the dissipative frictional contacts in the system. The ratio of the elastic to viscous scaling is tied to the nature of the frictional contact and a transition from adhesion- to load-controlled friction may occur in rougher samples. Our initial measurements point to such a transition, but additional measurements must be conducted. Since there is a sensitivity to the nature of the particle contacts, the influence of the wetting behavior, e.g. spreading coefficient and transition between Wenzel and Cassie-Baxter wetting regimes, should be systematically explored in this future work.

In this work, only pendular state capillary suspensions were investigated, but it is very likely that particle roughness also influences similar three and four phase systems that combine the themes of particles in contact and liquid wetting. Particle roughness will influence the detachment energy of particles adsorbed on small droplets in capillary state capillary suspensions.~\cite{Bossler2016} In four phase capillary foams with air as the extra phase, particle roughness changes the effective wettability and consequently the stability of these foams.~\cite{Zhang2017} Capillary condensation of a liquid on the particle surface is facilitated by particle roughness, which might affect the formation of liquid bridges in partially miscible capillary suspensions.~\cite{Fischer2021b} It would, therefore, be interesting to investigate the influence of roughness in these systems. For raspberry-like particles, the asperities protrude outwards, but particles containing pores have been shown to affect the morphology and rheology of capillary suspensions as well.~\cite{Bossler2016} Moreover, the detachment behavior of particles with chemically heterogeneous surface properties was found to be similar to that of physically rough particles, so this might also influence the yielding behavior of capillary suspensions.~\cite{Zanini2017} The frictional particle contacts, and thus particle roughness, are expected to be even more important for capillary suspensions with a higher volume fraction of solids.~\cite{Ahuja2017,Danov2018} Finally, the confocal microscopy and rheology measurements were performed separately in the present experiments. A logical continuation would be to perform simultaneous confocal and rheological measurements to more thoroughly investigate the yielding behavior of capillary suspensions with different particle and bridge shapes.   

\section*{Author Contributions}
JA and SB synthesized the particles and conducted the rheological and confocal experiments, MCRG and SDF helped with the AFM characterization of the rough particles, YZ helped with the synthesis and 3D-DLS measurements of the nanoparticles. EK supervised the research. JA wrote the original draft with reviewing and editing from EK.
\section*{Conflicts of interest}
The authors declare that they have no known competing financial interests or personal relationships that could have appeared to influence the work reported in this paper.
\section*{Acknowledgements}
The authors would like to thank financial support from the Research Foundation Flanders (FWO) Odysseus Program (grant agreement no. G0H9518N) and International Fine Particle Research Institute (IFPRI).

%\newpage

%%%END OF MAIN TEXT%%%

%%%REFERENCES%%%
%\A{Roughness} %You need to replace "rsc" on this line with the name of your .bib file
%\bibliographystyle{elsarticle-num} %the RSC's .bst file

\includepdf[pages=-]{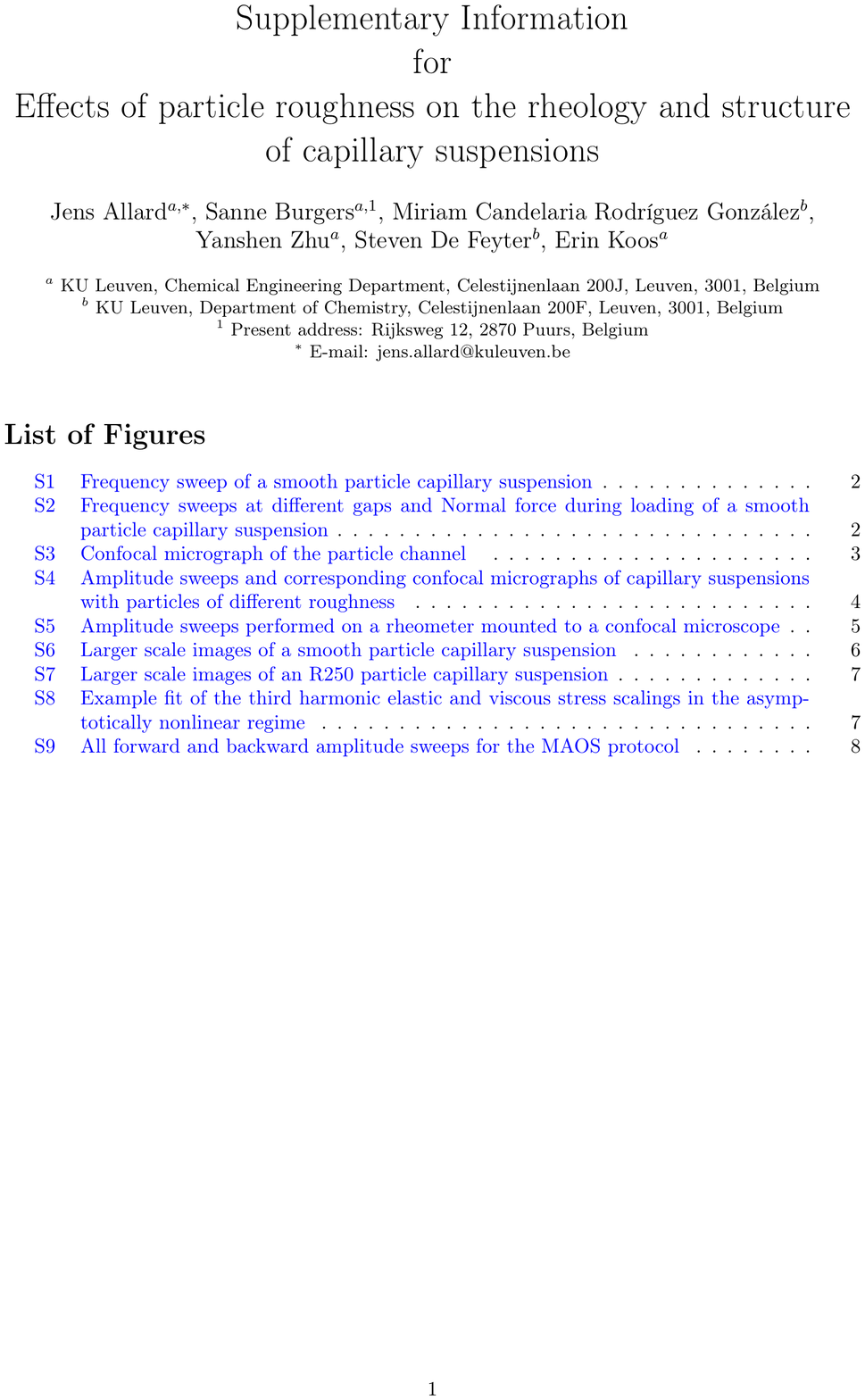}

\end{document}